\documentclass{emulateapj}
\usepackage{apjfonts}
\usepackage{epsf}
\usepackage{psfig}
\begin{document}

\slugcomment{Submitted to ApJL}
\shortauthors{Miller et al.}
\shorttitle{A Neutron Star Disk Wind}

\title{A Fast X-ray Disk Wind in the Transient Pulsar IGR
  J17480$-$2446 in Terzan 5}

\author{Jon~M.~Miller\altaffilmark{1},
        Dipankar~Maitra\altaffilmark{1},
        Edward~M.~Cackett\altaffilmark{2},
        Sudip~Bhattacharyya\altaffilmark{3},
        Tod.~E.~Strohmayer\altaffilmark{4}}

\altaffiltext{1}{Department of Astronomy, University of Michigan, 500
Church Street, Ann Arbor, MI 48109-1042, jonmm@umich.edu}

\altaffiltext{2}{Institute of Astronomy, University of Cambridge,
  Madingley Road, Cambridge, CB3 OHA, UK}

\altaffiltext{3}{Department of Astronomy and Astrophysics, Tata Institute of Fundamental Research, Mumbai 400005, India}

\altaffiltext{4}{Astrophysics Science Division, NASA Goddard Space Flight Center, Greenbelt, MD 20771}

\keywords{X-rays: binaries -- accretion, accretion disks -- stars: neutron -- pulsars: IGR J17480$-$2446}

\label{firstpage}

\begin{abstract}
Accretion disk winds are revealed in {\it Chandra} gratings spectra of
black holes.  The winds are hot and highly ionized (typically composed
of He-like and H-like charge states), and show modest blue-shifts.
Similar line spectra are sometimes seen in ``dipping'' low-mass X-ray
binaries, which are likely viewed edge-on; however, that absorption is
tied to structures in the outer disk, and blue-shifts are not
typically observed.  Here we report the detection of blue--shifted
He-like Fe XXV ($3100\pm 400$~km/s) and H-like Fe XXVI ($1000\pm
200$~km/s) absorption lines in a {\it Chandra}/HETG spectrum of the
transient pulsar and low-mass X-ray binary IGR J17480$-$2446 in Terzan
5.  These features indicate a disk wind with at least superficial
similarities to those observed in stellar-mass black holes.  The wind
does not vary strongly with numerous weak X-ray bursts or flares.  A
broad Fe K emission line is detected in the spectrum, and fits with
different line models suggest that the inner accretion disk in this
system may be truncated.  If the stellar magnetic field truncates the
disk, a field strength of $B = $0.7--4.0$\times 10^{9}~ {\rm G}$ is
implied, which is in line with estimates based on X-ray timing
techniques.  We discuss our findings in the context of accretion flows
onto neutron stars and stellar-mass black holes.
\end{abstract}

\section{Introduction}
IGR J17480$-$2446 was discovered on 10 October 2010 in an INTEGRAL
monitoring observation of the Galactic bulge (Bordas et al.\ 2010).
The source was found to be consistent with the position of the
globular cluster Terzan 5, which hosts the better-known transient and
Type-I X-ray burst source EXO 1745$-$248.  An affiliation with EXO
1745$-$248 was apparently strengthened with the detection of Type-I
bursts (Chevenez et al.\ 2010; Strohmayer \& Markwardt 2010).
Analysis of a prior {\it Chandra} image of Terzan 5, and additional
{\it Swift} observations revealed, however, that the source was not
EXO 1745$-$248, but rather a new transient source (Heinke et
al.\ 2010; also see Pooley et al.\ 2010).  The source was subsequently
given the name IGR J17480$-$2446, in recognition of its discovery with
INTEGRAL (Ferrigno et al.\ 2010).

Detailed timing analysis shows that IGR J17480$-$2446 is an 11~Hz
pulsar (Strohmayer \& Markwardt 2010).  It also appears that IGR
J17480$-$2446 evolved from an ``atoll'' into a ``Z'' source, based on
the path it traces in an X-ray color--color diagram and its rapid
variability components (including the possible detection of a kHz
quasi-periodic oscillation, or QPO, at 815~Hz; see Altamirano et
al.\ 2010).  Previously, only XTE J1701$-$462 had been observed to
evolve in this manner (Homan et al.\ 2010).  Slower, mHz QPOs were
also detected from IGR J17480$-$2446, caused by recurrent weak
bursts (Linares et al.\ 2010).  

Neutron stars in globular clusters are of particular importance
because the distance to globular clusters is often known precisely.
This can eliminate a major source of uncertainty when estimating a
blackbody emission radius.  The distance to Terzan 5 is likely 5.5~kpc
(Ortolani et al.\ 2007; also see Ransom 2007).  Motivated by the rare
opportunity to study a bursting neutron star in a globular cluster, we
proposed a {\it Chandra} Director's Time observation of IGR
J17480$-$2446.  

\section{Observation and Data Reduction}
IGR J17480$-$2446 was observed with {\it Chandra} on 24 October 2010
(ObsID 13161, SeqNum 401282), starting at 10:41:43 (UTC), for a total
exposure of 30.0~ksec.  The High Energy Transmission Gratings (HETG)
were used to disperse the incident flux onto the Advanced CCD Imaging
Spectrometer spectroscopic array (ACIS-S).  To prevent photon pile-up,
the ACIS-S array was operated in continuous clocking or ``GRADED\_CC''
mode, which reduced the nominal frame time from 3.2 seconds to 2.85
msec.  This mode is subject to calibration uncertainties, partially
owing to charge transfer inefficiency.  The zeroth order flux is
incident on the S3 chip, and frames from this chip can be lost from
the telemetry stream if a source is very bright.  We therefore used a
gray window over the zeroth order aimpoint; only one in 10 photons
were telemetered within this region.  For a longer discussion of this
mode, please see, e.g., Miller et al.\ (2006) and Miller et
al.\ (2008).

Data reduction was accomplished using CIAO version 4.1.  Time-averaged
first-order HEG and MEG spectra were extracted from the Level-2 event
file.  Redistribution matrix files (rmfs) were generated using the
tool ``mkgrmf''; ancillary response files (arfs) were generated
using ``mkgarf''.  The first-order HEG spectra and responses
were combined using the tool ``add\_grating\_orders''.  The spectra
were then grouped to require a minimum of 10 counts per bin.
Lightcurves of the dispersed photons were created using
``dmexract''.  Spectra from periods with and without bursts (see
below) were created by filtering the Level-2 event file using
``dmcopy'', specifying intervals derived using an independent
code.

All spectral analyses were conducted using XSPEC version 12.6.0.  All
errors quoted in this paper are 1$\sigma$ errors derived allowing all
model parameters to vary simultaneously.

%\begin{figure}
%\includegraphics[scale=0.50,angle=-90]{f1.ps}
\centerline{~\psfig{file=f1.ps,width=3.2in,angle=-90}~}
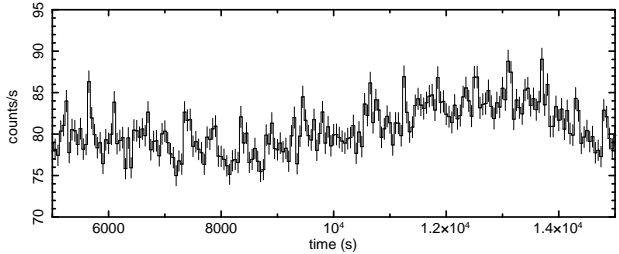
\figcaption[t]{\footnotesize A typical segment of the lightcurve of the
  dispersed spectrum is shown above, with 50s time bins.  The rapid
  flaring seen is likely muted bursting activity, and is the origin of
  the mHz QPOs detected in this source (see Linares et al.\ 2010).}
%\end{figure}
\medskip

\section{Analysis and Results}
\subsection{Bursts in the Lightcurve}
The full 0.3--10.0~keV lightcurve of IGR~J17480$-$2446 was searched
for bursts.  Figure 1 shows a segment of the total lightcurve; several
short, weak bursts are evident.  Chakraborty \& Bhattacharyya (2011)
suggest that the bursts seen in this transient are Type I bursts due
to nuclear burning on the stellar surface, not Type II bursts due to
an accretion disk process, as previously suggested by Galloway \& in
't Zand (2010).  Theoretical work shows that mHz QPOs can indeed be
due to marginally stable nuclear burning (Heger, Cumming, \& Woosley
2007).  We employed the search algorithm used by Maitra \& Miller
(2010) to flag bursts in the 0.3--10.0~keV lightcurve.  In total, 35
bursts were identified, generally confined within 100--200~s
intervals.  Time-selected spectra were made using the procedure
described above.

\subsection{The Spectral Continuum}
Preliminary fits to the spectra revealed a rather high column density,
approaching $10^{22}~ {\rm cm}^{-2}$.  For this reason, the MEG
spectra are not very sensitive to narrow atomic lines.  Moreover, the
flux calibration of the MEG in the Fe K band differs from the
better-calibrated HEG.  As a result, our spectral analysis focused
only on the summed first-order HEG spectrum.  Below 1.3 keV and above
9.0 keV, plausible continuum models leave residuals that are due to
calibration uncertainties.  More modest residuals are seen in the
1.7--3.0~keV band, likely owing to the complexity of modeling the
effective area of CCDs in the Si K band, the Ir coating on the {\it
  Chandra} mirrors, and chip gaps that fall in this range.  These
features have little impact on the goodness-of-fit statistics.  We
therefore fit the HEG spectra across the 1.3--9.0~keV band.

Although numerous bursts are evident in the source lightcurve, they
are short-lived features that represent a small fluctuations on top of
a high flux level.  Fits to the burst-free spectrum of IGR
J17480$-$2446 yield results that are consistent with fits to the
time-averaged spectrum; the 0.5--10~keV flux in the burst-free
spectrum is only 5\% lower than in the time-averaged spectrum.  (The
line properties measured during bursting phases, in the burst-free
spectrum, and the time-averaged spectrum, are also consistent.)  Owing
to this consistency, we focused on the time averaged spectrum of IGR
J17480$-$2446.

The continuum can be described well using a simple additive
model, consisting of a blackbody 

%\begin{figure}
%\includegraphics[scale=0.50,angle=-90]{f2.ps}
\centerline{~\psfig{file=f2.ps,width=3.2in,angle=-90}~}
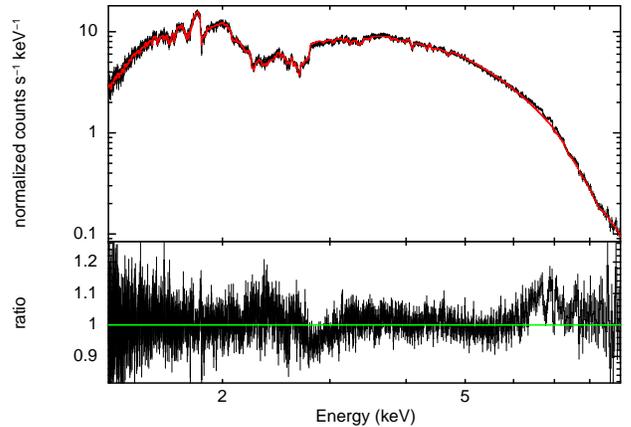
\figcaption[t]{\footnotesize The time-averaged {\it Chandra}/HETGS
  spectrum of IGR J17480$-$2446 is shown above, fitted with a
  simple blackbody plus power-law model.  Note the broad Fe K emission
  line, and Fe XXV and Fe XXVI absorption lines in the data/model ratio.}
%\end{figure}
\medskip

\noindent component and a power-law, both
modified by absorption along the line of sight.  In XSPEC parlance,
this model is ``tbabs*(bbodyrad$+$powerlaw)'', and it gives
$\chi^{2}/\nu = 3395.2/3255 = 1.04$ (where $\nu$ is the number of
degrees of freedom).  This model gives a column density of
${\rm N}_{\rm H} = 1.17\pm 0.04 \times 10^{22}~ {\rm cm}^{-2}$, a
blackbody temperature of kT$=1.18\pm 0.01$~keV, a blackbody
normalization of $267\pm 9$, a power-law index of $\Gamma = 1.26\pm
0.05$, and a power-law normalization of $0.46\pm 0.05$.  The
normalization of the ``bbodyrad'' model gives a color radius of
$9.0\pm 0.1$~km (${\rm K}_{\rm BB} = {\rm R}^{2}/{\rm D}^{2}_{10}$,
where ${\rm D}_{10}$ is the distance in units of 10~kpc); this value is
consistent with plausible stellar radii (e.g. Lattimer \&
Prakash 2006).  Extrapolating this simple model to the standard
0.5--10.0~keV soft X-ray band, it gives an unabsorbed flux of
$1.01(5)\times 10^{-8}~ {\rm erg}~ {\rm cm}^{-2}~ {\rm s}^{-1}$, which
corresponds to a luminosity of $3.7\pm 0.2 \times 10^{37}~ {\rm erg}~
{\rm s}^{-1}$.  Figure 2 shows the combined first-order HEG spectrum
of IGR J17480$-$2446, fitted with this continuum.

\subsection{The Line Spectra}
The spectrum of IGR J17480$-$2446 shows clear residuals in the Fe K
band (see Figures 2 and 3).  A broad emission line, whether a simple
Gaussian line or a relativistic line, gives an improvement to the
overall fit that is significant at far more than the $8\sigma$ level
of confidence (the F-statistic is $7.8\times 10^{-24}$).  To
characterize the absorption lines, we initially fit the broad line
using a simple Gaussian function.  The Gaussian model gives a centroid
energy of 6.76(3)~keV, which is broadly consistent with Fe XXV.  The
line strength is typical of neutron star LMXBs, with an equivalent
width of 112(9)~eV.  The width is measured to be 0.75(7)~keV (FWHM),
or roughly 0.1$c$.

Fits to the likely Fe XXV absorption line with a simple Gaussian give
a line centroid energy of 6.77(1) keV, indicating a blue-shift of
$3100 \pm 400$~km/s (for a rest energy of 6.700~keV; Verner, Verner,
\& Ferland 1996).  The Gaussian fit nominally indicates that the line
is resolved, with a width of $110\pm 20$~eV or $4800\pm 900$~km/s.  An
equivalent width of $12\pm 3$~eV is measured.  This line may be
saturated; its properties may not be accurately measured
using a Gaussian.

Fits to the likely Fe XXVI absorption line with a simple Gaussian give
a centroid energy of 6.988(4) keV, indicating a 

%\begin{figure}
%\includegraphics[scale=0.50,angle=-90]{f3.ps}
\centerline{~\psfig{file=f3.ps,width=3.2in,angle=-90}~}
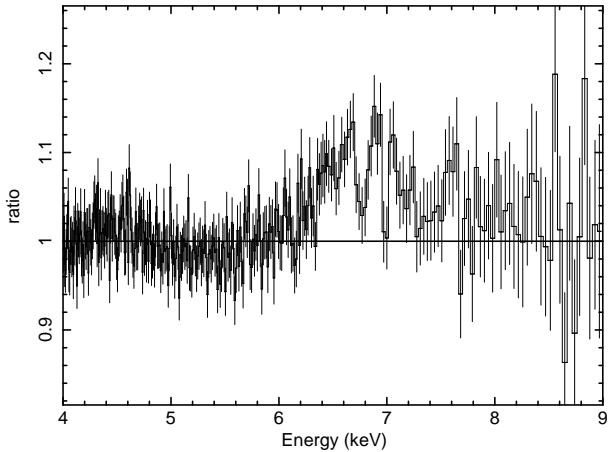
\figcaption[t]{\footnotesize This figure focuses on the Fe K range in
  the data/model ratio shown in Figure 2.  (The emission feature at
  7.6~keV is false and due to the addition of the first-order
  HEG spectra and responses.)}
%\end{figure}
\medskip

\noindent blue-shift of $1000 \pm 200$ km/s (for a rest energy of
6.966~keV; Verner, Verner, \& Ferland 1996).  As Figures 3 and 4
indicate, this line is quite narrow, and it is not resolved.  The 90\%
confidence upper limit on the width of this line is 13~eV or 600~km/s
(FWHM).  An equivalent width of $6 \pm 2$~eV is measured for this
line.  This phenomenological continuum and complex of three Fe lines
gives a very good fit: $\chi^{2}/\nu = 3248.3/3248.0$.  An apparent
feature remaining at 6.1~keV is not statistically significant.  The
inconsistency of the velocity shift in the He-like and H-like lines
could be real, but it may reflect possible saturation of the Fe XXV
line.  If it is real, the fact that the He-like line -- which likely
arises further from the ionizing flux than the H-like line -- has an
{\it higher} velocity, may signal that the wind is being accelerated.
Assuming that the observed absorption occurs on the linear part of the
curve of growth, the equivalent width of the Fe XXV line implies a
column of ${\rm N}_{Z} = 1.4(4) \times 10^{17}~ {\rm cm}^{-2}$; that
of the Fe XXVI line implies a column of ${\rm N}_{Z} = 1.3(4) \times
10^{17}~ {\rm cm}^{-2}$.

Modeling the broad feature with a relativistic line function
does not give a signicantly better fit,
compared to fits made with a simple Gaussian function.  Detecting
asymmetry in a broad line requires high sensitivity (see, e.g.,
Cackett et al.\ 2009, 2010; also see Miller et al.\ 2010).  Fits with
the ``diskline'' model give $r_{\rm in} = 20(2)~{\rm GM}/{\rm c}^{2}$,
or $41(5)$~km for a 1.4~${\rm M}_{\odot}$ neutron star.  This model
also gives $i = 18^{\circ}\pm 2^{\circ}$, ${\rm E} =
6.97_{-0.02}$~keV, and an equivalent width of $W = 130\pm 10$~eV.  The
line emissivity index was fixed at $q = -3$, and the outer line
emitting radius fixed at 1000~${\rm GM}/{\rm c}^2$.

We note that RXTE observed IGR J17480$-$2446 on 24
October 2010 (observation 95437-01-10-02).  The PCA spectra
appear to also reveal a broad emission line (Chakraborty \&
Bhattacharyya 2011).  Simple fits to the PCU2 spectrum (including
0.6\% systematic errors) in the 3--25~keV band strongly require a
line with a similar width ($0.9\pm 0.2$~keV, FWHM) and strength
(EW$=140\pm 20$~eV), but pegging at the lowest plausible energy for
Fe K (6.40~keV).  

The radius given by the relativistic line fit permits a constraint on
the stellar magnetic field strength.  Using equation (1) in Cackett et
al.\ (2009b), making the same assumptions regarding geometry and the
accretion efficiency, adopting an unabsorbed 0.1--30.0~keV flux of
$1.9\times 10^{-8}~ {\rm erg}~ {\rm cm}^{-2}~ {\rm s}^{-1}$ as the
bolometric flux, taking a distance of 5.5~kpc (Ortolani et a.\ 2007),
and assuming a mass of $1.4~{\rm M}_{\odot}$, we obtain $\mu = 4.5(8)\times
10^{26}~ {\rm G}~ {\rm cm}^{-3}$.  For a neutron star with a radius  

%\begin{figure}
%\includegraphics[scale=0.50,angle=-90]{f4.ps}
\centerline{~\psfig{file=f4.ps,width=3.2in,angle=-90}~}
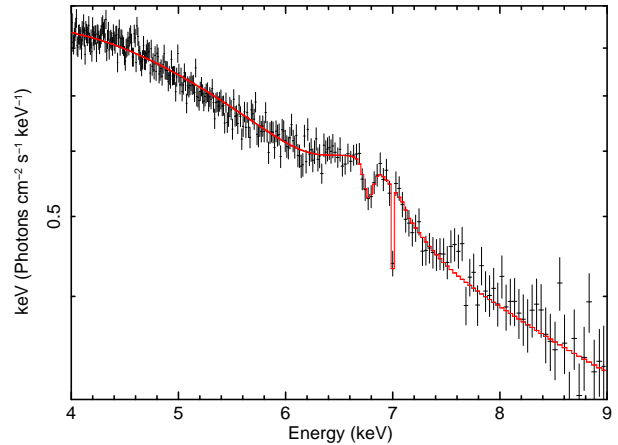
\figcaption[t]{\footnotesize The figure above is an ``unfolded'' or
  ``fluxed'' spectrum of IGR J17480$-$2446, based on phenomenological
  modeling to the spectrum in the Fe K band.  A broad Gaussian
  emission line and two narrow Gaussian absorption lines provide a
  good fit to the atomic features.}
%\end{figure}
\medskip

\noindent of 10~km, this gives a magnetic field strength of $B =
9(2)\times 10^{8} {\rm G}$ at the poles.  If a value of 0.5 (Long et
al.\ 2005) is assumed for the conversion factor from spherical to disk
accretion when balancing magnetic and ram pressures, the magnetic
field at the poles would be $B = 3(1)\times 10^{9}~ {\rm G}$.  Both
values are consistent with estimates based on X-ray timing (Papitto et
al.\ 2011).

\subsection{Photoionization Models}
In order to better model the absorption found in IGR J17480$-$2246, we
constructed a grid of multipicative models using XSTAR version 2.2
(Kallman \& Bautista 2001).  The blackbody plus power-law model
detailed above was extrapolated to the 0.1--30~keV band, and supplied
to XSTAR as the source of ionizing flux.  Solar abundances were
assumed.  The outflow must be equatorial, owing to the lack of narrow
emission lines in the spectra.  We therefore assumed a global covering
factor of 0.3.  Last, the grids considered here assumed a number
density of $n = 10^{12}~ {\rm cm}^{-3}$, consistent with
values used to characterize other winds in X-ray binaries.

Direct fits to the combined first-order HEG spectrum with the
multiplicative table model strongly suggest that the Fe XXV and XXVI
lines observed cannot arise in exactly the same gas.  Any single
absorption zone that fits the Fe XXV line well {\it also} predicts a
strong Fe XXVI line at a commensurate blue shift, which exceeds the
shift measured from the observed Fe XXVI line.  Similarly, any
absorption zone that fits the observed Fe XXVI line well,
under-predicts the Fe XXV line and fails to match its large observed
blue-shift.

A plausible fit to the spectrum ($\chi^{2}/\nu = 3306.1/3246$) can be
obtained using two absorption zones and a relativistic diskline model
for the broad emission line (see Figure 5).  The Fe XXV line is well
described by an absorber with ${\rm N}_{\rm H} = 3\times 10^{22}~
{\rm cm}^{-2}$, log($\xi$) = 3.0, at a blue shift of 3100~km/s.  The
Fe XXVI line is well described by a slightly more modest absorber: ${\rm
  N}_{\rm H} = 2\times 10^{22}~ {\rm cm}^{-2}$, log($\xi$) = 4.3, at
a blue shift of 500~km/s.  The higher velocity absorption zone
includes a strong Fe XXVI line that is not evident.  This may be
explained if the blue wing of a relativistic emission line (like that
described above) coincides with the Fe XXVI absorption.  The model
described here is not unique (for this reason, errors are not given),
and slightly different relativistic emission line profiles and
absorption zones can be combined to give similar outcomes.  

%\begin{figure}
%\includegraphics[scale=0.50,angle=-90]{f5.ps}
\centerline{~\psfig{file=f5.ps,width=3.2in,angle=-90}~}
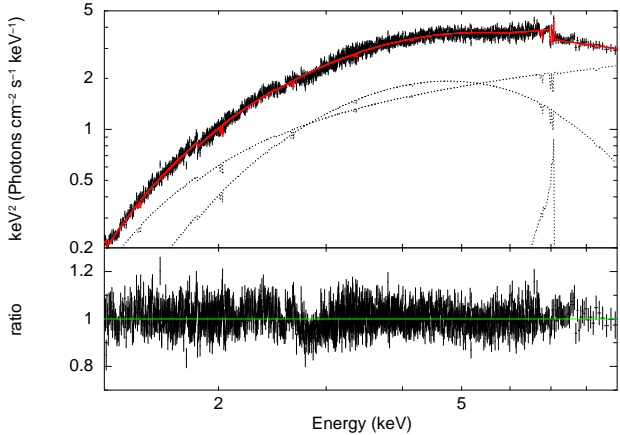
\figcaption[t]{\footnotesize The figure above shows unfolded spectrum
  of IGR J17480$-$2446, and the resulting data/model ratio.  The model
  used consists of a simple continuum, a relativistic disk line, and a
  two-zone ionized absorber generated using XSTAR.  The model does an
  excellent job of fitting the complex spectrum oberved in the Fe K
  band.}
%\end{figure}
\medskip

Assuming that the absorbing gas is a filled
volume of uniform density, the ionization parameter can be used to
derive a radius within which the wind must be launched (${\rm r}
\simeq {\rm L}/{\rm N}\xi$, where ${\rm L}$ is the source luminosity,
${\rm N}$ is the column density, and $\xi$ is the ionization
parameter).  Even the more highly ionized  absorber is relatively far
away from the central engine: ${\rm r} \simeq 3\times 10^{5}$~km, or
${\rm r} \simeq 1.5\times 10^{5}~ {\rm R}_{\rm Schw.}$ (assuming a
neutron star mass of $1.4 {\rm M}_{\odot}$).  This is within the
binary system of IGR 17480$-$2446: $a{\rm sin}(i) = 2.498(5)$~s, or
$7.5\times 10^{5}$~km (Papitto et al.\ 2001), and consistent with the
outer disk.

The mechanical luminosity in a wind is given by $L_{W} = 0.5 \dot{m}
v^{2}$.  In practice, it is necessary to rewrite this
equation in terms of the ionization parameter and a geometric factor:
$L_{W} = 0.5 m_{p} \Omega v^{3} (L/\xi)~$, where $m_{p}$ is the mass
of the proton, $\Omega$ is the fraction of $4\pi$ covered by the wind
($\Omega = 0.3$ is assumed in our XSTAR grids), $v$ is the wind
velocity, $L$ is the ionizing luminosity, and $\xi$ is the ionization
parameter.  The faster, less ionized absorption zone gives a
mechanical luminosity of $L_{W} \simeq 2.8\times 10^{35}~ {\rm erg}~
{\rm s}^{-1}$, or $L_{W}/L_{accr.} \simeq 8\times 10^{-3}$.

\section{Discussion and Conclusions}
We have analyzed a {\it Chandra}/HETG observation of the transient
X-ray pulsar and low-mass X-ray binary IGR J17480$-$2446 in Terzan 5.
The Fe K band can be decomposed into a broad emission line consistent
with He-like emission excited at inner edge of a radially-truncated
accretion disk, and blue-shifted He-like and H-like Fe absorption
lines that likely arise in an X-ray disk wind.  The implications of a
truncated inner accretion disk in IGR J17480$-$2446 and a disk wind
are of particular interest, and they are explored in this section.

At 11~Hz, IGR J17480$-$2446 is a much slower pulsar than e.g. SAX
J1808.4$-$3658, which has a frequency of 401~Hz (Wijnands \& van der
Klis 1998), and in which relativistic line spectroscopy and X-ray
timing both imply a smaller inner disk radius and smaller stellar
magnetic field (Cackett et al.\ 2009b; Papitto et al.\ 2009, Hartman
et al.\ 2008).  In SAX J1808.4$-$3658, then, it is not surprising that
the source shows kHz QPOs, which are often associated with orbits in
the inner disk, and connections between the disk and the star.  

At a flux level similar to that measured in this {\it
  Chandra} observation (after accounting for a distance disparity),
Altimirano et al.\ (2010) reported the detection of QPOs at 48~Hz,
173~Hz, and 815~Hz.  If the fastest QPO signals a Keplerian orbital
frequency, a radius of approximately $9~ {\rm GM}/{\rm c}^{2}$ is
implied.  If the inner disk was still truncated at $20(2)~ {\rm
  GM}/{\rm c}^{2}$, then the kHz QPOs would have to be produced
interior to the disk truncation radius.  Fourier-resolved spectroscopy
has shown that the variable part of such spectra may originates in
the boundary layer (Revnivtsev \& Gilfanov 2006); thus, it is possible
that kHz QPOS originate in the boundary layer.  A more likely
explanation is that the disk simply extended closer to the neutron
star when kHz QPOs were detected.

The disk wind found in IGR J17480$-$2446 may be the clearest detection
of such a flow in a neutron star system.  Iron absorption lines from
highly ionized gas are clearly detected in a number of ``dipping''
neutron star systems, but this absorption is not typically
blue-shifted, and likely occurs in the outer accretion disk of these
nearly edge-on sources (see, e.g., Diaz Trigo et al.\ 2006).  An
ionized outflow is clearly detected in the persistent spectrum of GX
13$+$1 (Ueda et al.\ 2004); however, this source shows frequent dips,
and this casts some doubt on the nature of the absorption.  Similarly,
an ionized outflow is detected in Circinus X-1 (Brandt \& Schulz
2000), but it is difficult to rule out a massive companion wind as the
source of the outflow.  An ionized X-ray wind was
detected in a {\it Chandra} spectrum of the pulsar 1A 0535$+$262
(Reynolds \& Miller 2010).  The flow is likely a disk wind, but it is
difficult to entirely rule out absorption in the wind of the massive B
companion star in 1A 0535$+$262.

It is interesting to explore why disk winds are not regularly detected
in the relatively large sample of persistent ``Z'' and ``atoll''
neutron stars.  A recent survey of {\it Chandra}/HETG spectra does not
find evidence for ionized outflows (Cackett et al.\ 2009); in some
cases the sensitivity of the spectra may have simply been
insufficient.  However, even in deep observations of Cygnus X-2,
blue-shifted absorption is not reported in the non-dip spectra (Schulz
et al.\ 2009). 

An intriguing possibility is that the outflow observed in IGR
J17480$-$2446 may be a different kind of disk wind than those observed
in stellar-mass black holes.  The winds seen in stellar-mass black
holes originate close to the black hole (e.g. within $1000~GM/c^{2}$;
see Miller et al.\ 2006a, 2006b, 2008; Kubota et al.\ 2007; Neilsen \&
Lee 2009), and can carry away a good fraction of the accreting gas.
Thus, the winds in stellar-mass black holes may be at least partially
driven by magnetic pressure (see Miller et al.\ 2008).  In contrast,
X-ray absorption in the disk wind found in IGR J17480$-$2446
originates two orders of magnitude further from the central engine
(measured in units of $GM/c^{2}$).  The high ionization parameter of
the wind in IGR J17480$-$2446 makes it unlikely that radiation
pressure can drive the wind (e.g. Proga et al.\ 2000), but Compton
heating of the outer disk could plausibly drive such a wind
(e.g. Begelman, McKee, \& Shields 1983).

\vspace{0.2in}
We thank Harvey Tananbaum, Belinda Wilkes, and Andrea Prestwich for
executing this observation.  We acknowledge David Pooley and Jeroen
Homan for helping to coordinate different observations of IGR
J17480$-$2446.  We thank Tim Kallman and Cole Miller for helpful
discussions.  Finally, we thank the anonymous referee for a helpful
review.  JMM ackhowledges support from the {\it Chandra} Guest
Observer program.

%----------------------------------------------------------------------

\end{document}